\documentclass [11pt, letterpaper]{article}
\pdfoutput=1 

\usepackage{fullpage}
\usepackage{amsmath}
\usepackage{amssymb}
\usepackage{graphicx}
\usepackage{mathrsfs}
\usepackage{wrapfig}
\usepackage{boxedminipage}
\usepackage{epsfig}
\usepackage{setspace}
\usepackage{subfigure}
\usepackage{float}
\usepackage{hyperref}
\usepackage{enumerate}
\usepackage{cite}
\usepackage[font=footnotesize,labelfont=rm]{caption}

\begin{document}

\begin{flushright}
{\tt arXiv:1512.01746}
\end{flushright}

{\flushleft\vskip-1.35cm\vbox{\includegraphics[width=1.25in]{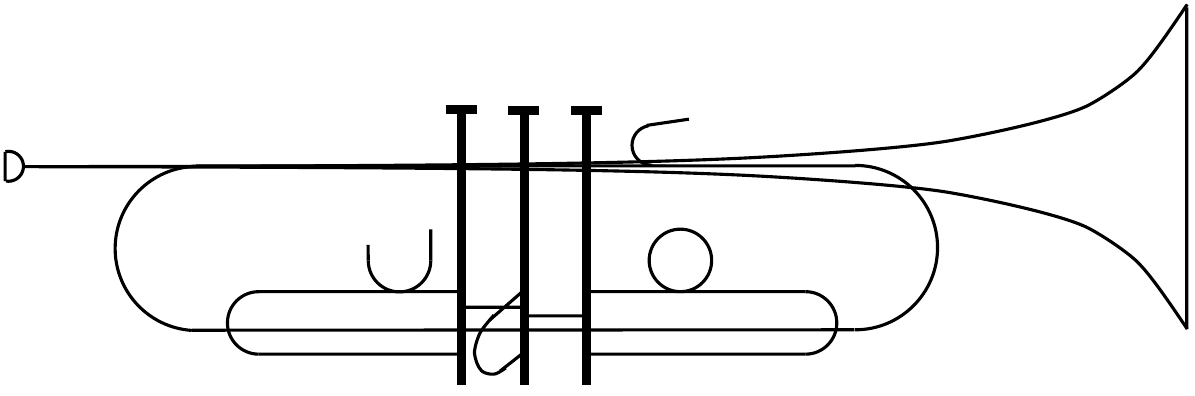}}}

\bigskip
\bigskip
\bigskip
\bigskip

\bigskip
\bigskip
\bigskip
\bigskip

\begin{center} 

{\Large\bf  Born--Infeld AdS Black Holes as Heat Engines}





\end{center}

\bigskip \bigskip \bigskip\bigskip

\centerline{\bf Clifford V. Johnson}

\bigskip
\bigskip
\bigskip
\bigskip

  \centerline{\it Department of Physics and Astronomy }
\centerline{\it University of
Southern California}
\centerline{\it Los Angeles, CA 90089-0484, U.S.A.}

\bigskip

\centerline{\small \tt johnson1,  [at] usc.edu}

\bigskip
\bigskip


\begin{abstract} 
\noindent  We study the efficiency of heat engines that perform mechanical work {\it via} the $pdV$ terms   present in the First Law in extended gravitational thermodynamics. We use charged black holes as  the working substance, for a particular choice of engine cycle. The context is Einstein  gravity with negative cosmological constant and a Born--Infeld non--linear electrodynamics sector. We compare the results for these ``holographic'' heat engines to previous results obtained for Einstein--Maxwell black holes, and for the case where there is a  Gauss--Bonnet sector. 
\end{abstract}
\newpage \baselineskip=18pt \setcounter{footnote}{0}

\section{Introduction}
%
This paper presents a class of corrections to the efficiency of holographic heat engines which have charged black holes as the working substance. The corrections arise as a result of non--linear extension of the electromagnetic sector, the Born--Infeld action. Holographic heat engines were defined in ref.\cite{Johnson:2014yja}.  They are a natural concept in {\it extended}  gravitational thermodynamics, which, in making the cosmological constant ($\Lambda$) dynamical in a theory of gravity, supplies\footnote{Here we are using geometrical units where $G,c,\hbar,k_{\rm B}$ have been set to unity. 
} a pressure variable $p=-\Lambda/8\pi$ and its conjugate volume~$V$ (see  refs.\cite{Kastor:2009wy,Caldarelli:1999xj,Wang:2006eb,Sekiwa:2006qj,LarranagaRubio:2007ut,Dolan:2010ha,Cvetic:2010jb,Dolan:2011jm,Dolan:2011xt,Henneaux:1984ji,Teitelboim:1985dp,Henneaux:1989zc}). One may extract mechanical work {\it via} the  $pdV$ term in the First Law of Thermodynamics, and so it is possible to define a cycle in state space during which there is a net input heat $Q_H$ flow, a net output flow $Q_C$ and a net output work $W$, such that $Q_H=W+Q_C$. The efficiency is then $\eta=W/Q_H$. Its value is determined by the equation of state of the system  and the choice of cycle in state space. The gravitational solution (a black hole, in our case) supplies the equation of state: The temperature $T$, entropy $S$, enthalpy~$H$ and other quantities can be defined\cite{Bekenstein:1973ur,Bekenstein:1974ax,Hawking:1974sw,Hawking:1976de,Kastor:2009wy}, and there are relations between them. 
 \begin{wrapfigure}{l}{0.43\textwidth}
{\centering
\includegraphics[width=2.3in]{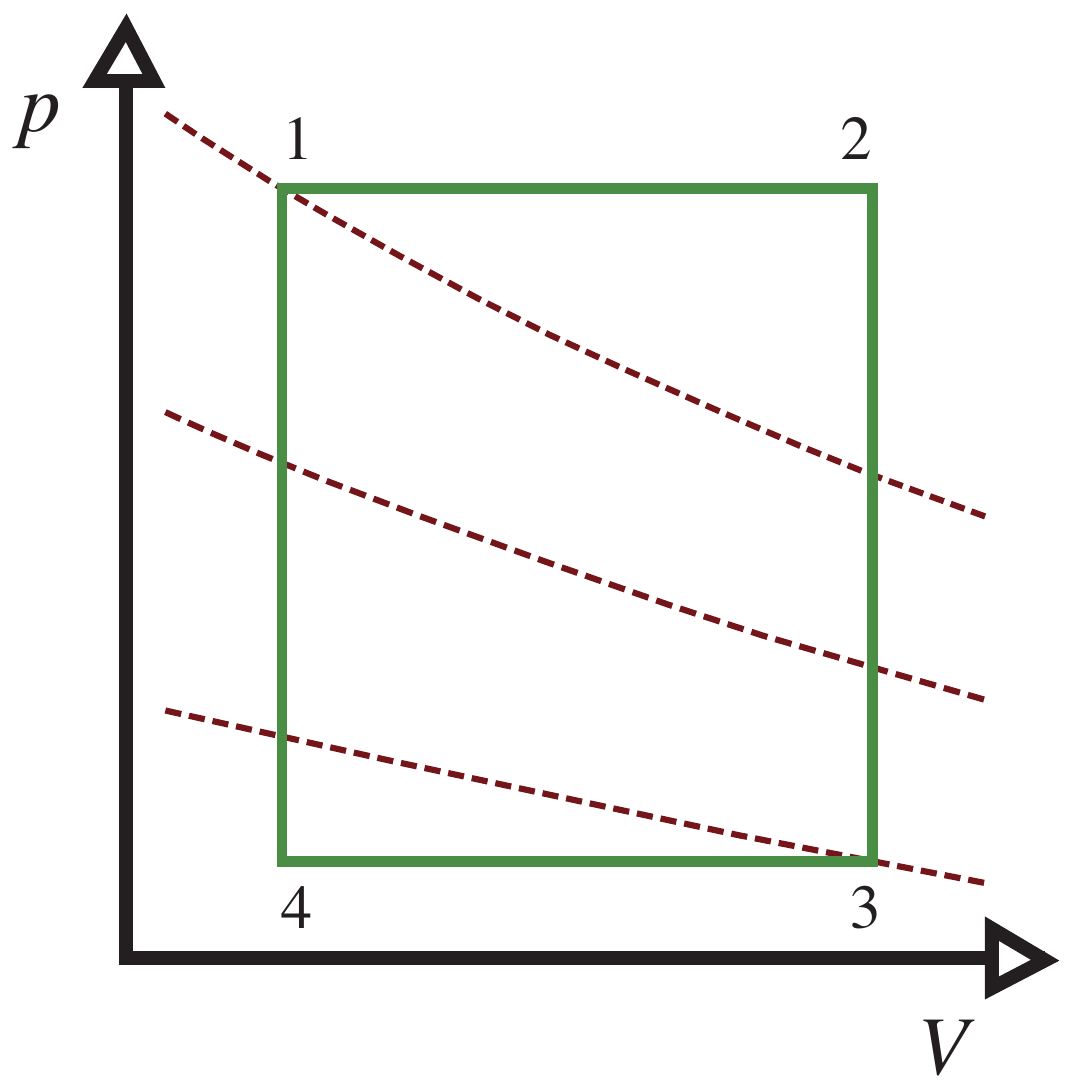} 
   \caption{\footnotesize  Our engine.}   \label{fig:cyclesb}
}
\end{wrapfigure}
We will choose the cycle given in figure~\ref{fig:cyclesb}, following earlier work in refs.\cite{Johnson:2014yja,Johnson:2015ekr}, where it  is explained why that is a natural choice for  static black holes, which  we will study here\footnote{Refs.\cite{Belhaj:2015hha,Sadeghi:2015ksa,Caceres:2015vsa,Setare:2015yra} have since done further studies of such heat engines. }.

The Born--Infeld action\cite{Born:1933,Born:1934ji,Born:1934gh} is a non--linear generalization of the Maxwell action\footnote{Here, we follow a common strand of terminology in the literature: Strictly speaking, the displayed action~(\ref{eq:born-infeld})  is due to Born\cite{Born:1933,Born:1934ji}. In the original $D=4$ context, the full Born--Infeld action has under the square root a quartic  term $(\mathbf{B}\cdot\mathbf{E})^2$  as well as the quadratic term $\mathbf{E}^2-\mathbf{B}^2$  shown.  However, in the case of vanishing magnetic sector (as will be true in this paper), the actions  have the same content. The   action~(\ref{eq:born-infeld}),  in any number of dimensions, is often referred to as Born--Infeld in the literature --- as is the $D$--dimensional version of the full action that Born and Infeld wrote\cite{Born:1934gh}  (with ${\rm det}(\eta_{\mu\nu}+F_{\mu\nu}/\beta)$ under the square root). Thanks to David Chow for prompting the clarification after an earlier version of this manuscript appeared.}, controlled by a parameter $\beta$:
\begin{equation}
{\cal L} (F)=4\beta^2\left(1-\sqrt{1+\frac{F^{\mu\nu}F_{\mu\nu}}{2\beta^2}}\right)\ ,
\label{eq:born-infeld}
\end{equation}
where  in the limit $\beta\to\infty$ we recover the Maxwell action. The completion of Maxwell into Born--Infeld is quite natural in string theory\cite{Gibbons:2001gy}, where $1/\beta \propto 2\pi\alpha^\prime$, {\it i.e.} it is like the tension of the string. The corrections about the $\beta\to\infty$ expansion (the famous zero--slope limit) is an infinite family of $\alpha^\prime$ corrections. In studying the efficiency of our black hole heat engine as a function of $\beta$, we can therefore think of it as capturing the effects of $\alpha^\prime$--like corrections in some underlying string theory model, but it is not necessary to do so. It is interesting enough to consider the system  in its own right.

Another possible context for this study is the fact that we will work in negative cosmological constant (defining a positive pressure), for which such physics has an holographic duality\cite{Maldacena:1997re,Witten:1998qj,Gubser:1998bc,Witten:1998zw,Aharony:1999t} to non--gravitational field theories in one dimension fewer, at large $N$ (where $N$ is the rank of a field theory gauge group, or an analogue thereof). As pointed out in ref.\cite{Johnson:2014yja}, since changing $\Lambda$ involves changing the $N$ (or analogues thereof) of the dual theory, the heat engine cycle is a kind of tour on the {\it space} of field theories rather than staying within one particular field theory\footnote{As pointed out in ref.\cite{Karch:2015rpa} it also involves changing the size of the space the field theories live on.}. Our studies therefore concern corrections to the physics of such tours as well, but our focus here will  be to study the properties of our new black hole engines in Born--Infeld for their own sake, leaving  examination of the implications for  those possible applications for another time.

Once we have extracted the efficiency of our engines in the presence of Born--Infeld (and we will do so working in a high temperature limit) we will compare the results to the Einstein--Maxwell case, and also contrast it with the results obtained in ref.\cite{Johnson:2015ekr} for another class of $\alpha^\prime$--like corrections, the Gauss--Bonnet case. In this way we'll have studied two distinct classes of corrections to these engines' efficiency:  Corrections to the Einstein sector, and corrections to the Maxwell sector.

\section{The Black Holes and the Equation of State}
\label{sec:example}
Our  Einstein--Hilbert--Born--Infeld bulk action in $D$--dimensions is:
\begin{equation}
I=\frac{1}{16\pi }\int \! d^Dx \sqrt{-g} \left(R-2\Lambda +{\cal L}(F)\right)\ ,
\label{eq:action}
\end{equation}
with ${\cal L}(F)$ given in equation~(\ref{eq:born-infeld}). The cosmological constant  sets a length scale $l$ according to: 
\begin{equation}
\Lambda=-\frac{(D-1)(D-2)}{2l^2}\ .
\label{eq:cosmocon}
\end{equation}
 The black hole has mass and charge  parameters $m$ and ${q}$, with metric\cite{Fernando:2003tz,Cai:2004eh,Dey:2004yt} 
\begin{equation}
ds^2 = -Y( r)dt^2
+ {dr^2\over Y(r)} + r^2 d\Omega^2_{D-2} \ , 
\label{eq:staticform}
\end{equation}
where $d\Omega^2_{D-2}$ is the metric on a round $D-2$ sphere with volume $\omega_{D-2}$, and
\begin{eqnarray}
&&Y( r) = 1-\frac{m}{r^{D-3}}+\frac{r^2}{l^2}+
\frac{4\beta^2r^2}{(D-1)(D-2)}\left(1-\sqrt{1+\frac{(D-2)(D-3)q^2}{2\beta^2 r^{2D-4}}}\right)\nonumber\\
&&\hskip 2cm+\frac{2(D-2)q^2}{(D-1)r^{2D-4}}\,\,{}_2F_1\left[\frac{D-3}{2D-4},\frac12,\frac{3D-7}{2D-4},-\frac{(D-2)(D-3)q^2}{2\beta^2 r^{2D-4}}\right]  
\ ,
\end{eqnarray}
where ${}_2F_1$  is the hypergeometric function. The  gauge potential  is:
\begin{equation}
A_t = -\frac{q}{c}\frac{1}{r^{D-3}} {}_2F_1\left[\frac{D-3}{2D-4},\frac12,\frac{3D-7}{2D-4},-\frac{(D-2)(D-3)q^2}{2\beta^2 r^{2D-4}}\right] \ , \quad {\rm with}\quad c=\sqrt{\frac{2(D-3)}{D-2}}\ .
\label{eq:gaugepotential}
\end{equation}
The mass and charge of the solution are given by:
\begin{equation}
M=\frac{(D-2)\omega_{D-2}}{16\pi} m \, \quad{\rm and}\quad Q=\sqrt{2(D-2)(D-3)}\left(\frac{\omega_{D-2}}{8\pi}\right) q \ .
\label{eq:paramters}
\end{equation}
\noindent
Given an horizon of radius $r_+$ (the largest root of $Y(r_+)=0$), the   temperature $T$, entropy $S$, and volume $V$ are given by\cite{Cai:2004eh,Dey:2004yt,Zou:2013owa}:
\begin{eqnarray}
 \label{eq:teepee}
 &&T=
 \frac{1}{4\pi}\left(16\pi p \frac{r_+}{(D-2)}+\frac{(D-3)}{r_+}+\frac{4\beta^2r_+^2}{(D-2)}\left(1-\sqrt{1+\frac{(D-2)(D-3)q^2}{2\beta^2 r_+^{2D-4}}}\right)\right)\ ,
\\ 
&& S=\frac{\omega_{D-2}}{4}r_+^{D-2}\ ,\quad  {\rm and}\quad V=\frac{\omega_{D-2}}{(D-1)} r_+^{D-1}\ ,
 \end{eqnarray}
 where we have used that $p=-\Lambda/8\pi$ and equation~(\ref{eq:cosmocon}). The temperature expression~(\ref{eq:teepee}) can be re--arranged into an equation of state (actually a family of equations of state parameterized by $q$, which we will keep fixed) in the $p{-}r_+$ plane, or equivalently  the $p{-}V$ plane.
As noted in ref.\cite{Johnson:2015ekr}, in the high temperature limit, the leading behaviour of this equation of state is:
\begin{equation}
p V^{1/(D-1)}\sim T\ ,
\label{eq:idealgaslaw}
\end{equation}
 a sort of ideal gas limit for our black holes. At lower temperatures there can be quite non--trivial behaviour as a coming from multivaluedness of the state curve (generalizing what was found for Reissner--Nordstr\"om black holes in refs.\cite{Chamblin:1999tk,Chamblin:1999hg,Kubiznak:2012wp}) giving rise to non--trivial phase transitions\cite{Fernando:2006gh,Myung:2008eb,Gunasekaran:2012dq,Banerjee:2012zm}, which  will not be our focus here.

 \section{The Engine Efficiency}

 \subsection{The Specific Heat}
 \label{sec:specificheat1}
 The specific heat at constant pressure   $C_p\equiv T\partial S/\partial T|_{p}$ is the next quantity we need. It is most easy to compute it in terms of $r_+$, as discussed in ref.\cite{Johnson:2015ekr}, and the result  is, for general $D$:
 \begin{eqnarray}
  \label{eq:veryspecific}
 &&C_p=\frac{(D-2)\omega_{D-2}}{4}r_+^{D-2}\times \\
&& \hskip1.5cm \left(\frac{\frac{16\pi }{(D-2)(D-3) } p\, r_+^{2D-4}+{r_+^{2D-6}}+\frac{4\beta^2 }{(D-2)(D-3) }\left(1-{\cal R}^\frac12\right)r_+^{2D-4}}{\frac{16\pi }{(D-2)(D-3)} p\,r_+^{2D-4}-{r_+^{2D-6}}+\frac{4\beta^2 }{(D-2)(D-3) }\left(1-{\cal R}^\frac12 \right)r_+^{2D-4}+2(D-2)q^2{\cal R}^{-\frac12}}\right) \ , \nonumber
 \end{eqnarray}
 where
 \begin{equation}
 {\cal R}\equiv  1+\frac{(D-2)(D-3)q^2}{2\beta^2 r_+^{2D-4}}\ .
 \end{equation}

\subsection{The  Efficiency $\eta$}
For our engine cycle defined in figure~\ref{fig:cyclesb}, we have 
\begin{equation}
 W= \left(V_2-V_1\right)(p_1-p_4)\ ,
 \label{eq:nicework}
 \end{equation}
 where the subscripts refer to the quantities evaluated at the corners labeled (1,2,3,4). The heat flows take place  along the top and bottom, with the upper isobar  giving the  net inflow of heat:
 \begin{equation}
 Q_H=\int_{T_1}^{T_2} C_p(p_1,T) dT\ ,
 \label{eq:hothothot}
\end{equation}
 The efficiency is then $\eta=W/Q_H$. 
 
 \subsection{The High Temperature Limit}
  \label{sec:high-temperature}
 To get an explicit expression for $C_p$ in terms of $T$ so that we can integrate along the isobar, we  take a high temperature limit,  solving for $r_+$ perturbatively in a large $T$ expansion, using equation~(\ref{eq:teepee}), and substituting into~(\ref{eq:veryspecific}). We expand $V$ in the same way. For example, in $D=4$:
\begin{eqnarray}
&&r_+=\frac{1}{2}\frac{T}{p}-\frac{1}{4\pi T}+\frac18\frac{p(8\pi p q^2-1)}{\pi^2 T^3}+\frac18\,{\frac {{p}^{2}
 \left( 16\,{q}^{2}p\pi -1 \right) }{{\pi }^{3}{T}^{5}}} \nonumber\\
 &&\hskip4cm+\left( -\frac{1}{32}\,{\frac {{p}^{3} \left( 5-120\,{q}^{2}p\pi +192\,{q}^{4}{p}^{2}{
\pi }^{2} \right) }{{\pi }^{4}}}-4\,{\frac {{p}^{6}{q}^{4}}{\pi \,{
\beta}^{2}}} \right) \frac{1}{T^7}
+\cdots\ , \nonumber \\
&&V=\frac{4\pi}{3}r_+^3= \frac{\pi}{6p^3}T^3-\frac14 \frac{T}{p^2}+\frac{q^2}{T}
+\frac{1}{48}\,{\frac {(48\,{q}^{2}p\pi -1)}{{\pi }^{2}{T}^{
3}}}\nonumber\\
&&\hskip4cm-\frac{1}{32}\,{\frac {p \left( 1-48\,{q}^{2}p\pi +
128\,{q}^{4}{p}^{2}{\pi }^{2}+128\,{p}^{3}{q}^{4}{\pi }^{3}/{\beta}^{2}
 \right) }{{\pi }^{3}{T}^{5}}}
+\cdots\ ,\nonumber \\
&& \int \!C_pdT=\frac{\pi}{6p^2}T^3+\frac18\frac{(16\pi pq^2-1)}{\pi T}
+\frac{1}{12}\,{\frac {p \left( 24\,{q}^{2}p\pi -1 \right) }{{\pi 
}^{2}{T}^{3}}}\nonumber\\
&&\hskip4cm-\frac15\, \left( {\frac {3}{32}}\,{\frac {{p}^{2} \left( 5-
160\,{q}^{2}p\pi +320\,{q}^{4}{p}^{2}{\pi }^{2} \right) }{{\pi }^{3}}}
+24\,{\frac {{p}^{5}{q}^{4}}{{\beta}^{2}}} \right) \frac{1}{T^5}
+\cdots
\label{eq:4dexpansions}
\end{eqnarray} 
and in $D=5$:
\begin{eqnarray}
&&r_+=\frac{3}{4}\frac{T}{p}-\frac{1}{2\pi T}-\frac{p}{3\pi^2 T^3}+\frac{4}{81}\frac{p^2(32q^2\pi^2p^2-9)}{\pi^3 T^5}
\nonumber\\
&&\hskip1.5cm
+{\frac {4}{81}}\,{\frac 
{{p}^{3} \left( 128\,{q}^{2}{p}^{2}{\pi }^{2}-15 \right) }{{\pi }^{4}{
T}^{7}}}+{\frac {112}{729}}\,{\frac {{p}^{4} \left( 128\,{q}^{2}{p}
^{2}{\pi }^{2} -9\right) }{{\pi }^{5}{T}^{9}}}\nonumber\\
&&\hskip1.5cm+ \left( -{\frac {32}{
19683}}\,{\frac {{p}^{5} \left( -34560\,{q}^{2}{p}^{2}{\pi }^{2}+10240
\,{q}^{4}{p}^{4}{\pi }^{4}+1701 \right) }{{\pi }^{6}}}-{\frac {131072}
{19683}}\,{\frac {{p}^{10}{q}^{4}}{\pi \,{\beta}^{2}}} \right) \frac{1}{T^{
11}}
+\cdots\ , \nonumber\\
\end{eqnarray}
with
\begin{eqnarray}
&&V=\frac{\pi^2}{2}r_+^4= \frac{81}{512}\frac{\pi^2}{p^4}T^4-\frac{27}{64} \frac{\pi T^2}{p^3}+\frac{9}{64 p^2}+\frac43\frac{\pi p q^2}{T^2}
\nonumber\\
&&\hskip1.5cm +{\frac {1}{96}}\,{\frac {(256\,{q}^{2}
{p}^{2}{\pi }^{2}-3)}{{\pi }^{2}{T}^{4}}}+{\frac {1}{108}}\,{\frac {p
 \left( 640\,{q}^{2}{p}^{2}{\pi }^{2} -9\right) }{{\pi }^{3}{T}^{6}}}\nonumber\\
&&\hskip1.5cm-{\frac {1}{2916}}\,{\frac {{p}^{2} \left( -40320\,{\pi }^{
2}{p}^{2}{q}^{2}+28672\,{\pi }^{4}{p}^{4}{q}^{4}+567+16384\,{q}^{4}{p}^{5}{\pi }^{5}/ {\beta}^{2}\right) }{{\pi }
^{4}{T}^{8}}}
+\cdots\ ,\nonumber \\ 
&&\int \!C_pdT=\frac{81}{512}\frac{\pi^2}{p^3}T^4-\frac{27}{128} \frac{\pi T^2}{p^2} +\frac{1}{96}\frac{(192\pi^2 p^2q^2-9)}{\pi T^2}
\nonumber\\
&&\hskip1.5cm+{\frac {5}{288}}\,{\frac {p \left( 
256\,{q}^{2}{p}^{2}{\pi }^{2} -9\right) }{{\pi }^{2}{T}^{4}}}+{\frac {7}
{216}}\,{\frac {{p}^{2} \left( 320\,{q}^{2}{p}^{2}{\pi }^{2}-9
 \right) }{{\pi }^{3}{T}^{6}}}\nonumber\\
&&\hskip1.5cm-{\frac {1}{324}}\,{\frac {{p}^{3}
 \left( -8064{\pi }^{2}{p}^{2}{q}^{2}+4096{
\pi }^{4}{p}^{4}{q}^{4}+189 +2048\,{q}^{4}{p}^{5}{\pi }^{5}/\beta^2\right) }{{\pi }^{4}{T}^{8}}}
+\cdots
\label{eq:5dexpansions}
\end{eqnarray} 
These expressions for $V$ and $\int C_p dT$ can now be used to compute the efficiency {\it via} equations~(\ref{eq:nicework}) and~(\ref{eq:hothothot}),  taking their ratio. Both quantities are evaluated at the pressure of the upper isobar,~$p_1$, and because $pV$ and $\int C_p dT$  have identical leading terms, we have $\eta=(1-p_4/p_1)+\cdots$ with corrections that can be evaluated readily by substitution (see refs.\cite{Johnson:2014yja,Johnson:2015ekr} for further discussion). 

A striking feature of  these expansions is how late $\beta$ enters: It is at order $T^{-5}$ for  $D=4$ and at order $T^{-8}$ for $D=5$. This will mean (as we shall see in the next section) that the variation in the efficiency as a function of  $\beta$ will be  somewhat understated as compared to  the variation with $\alpha$ (the coefficient of the Gauss--Bonnet action)  in  the companion study of ref.\cite{Johnson:2015ekr}. There, $\alpha$ appears  immediately at the next--to--leading term at order $T^2$.

\section{Two Studies of $\eta(\beta)$}
\label{sec:twostudies}
Equipped with our high temperature expansion, we can now study  the efficiency as a function of~$\beta$, seeing how $\eta(\beta)$ behaves as we move away the $\beta=\infty$ (Maxwell) limit. The  efficiency in the Einstein--Maxwell limit will be  denoted $\eta^{\phantom{9}}_0=\lim_{\beta\to\infty}\eta(\beta)$. We will study  the two schemes that were defined in ref.\cite{Johnson:2015ekr}, determined by   what parameters of the cycle we specify and  hold fixed as we  change $\beta$.

\subsection{Scheme 1}
Here, for our engine cycle (see figure~\ref{fig:cyclesb}) we  specify the two operating pressures  $(p_1,p_4)$ and the two temperatures $(T_1,T_2)$. 
We can evaluate the efficiency in this scheme as a function of $\beta$, seeing how it moves away from the benchmark  
$\eta^{\phantom{9}}_0$    of   Maxwell electrodynamics.

 \begin{figure}[h]
{\centering
\subfigure[]{\includegraphics[width=2.6in]{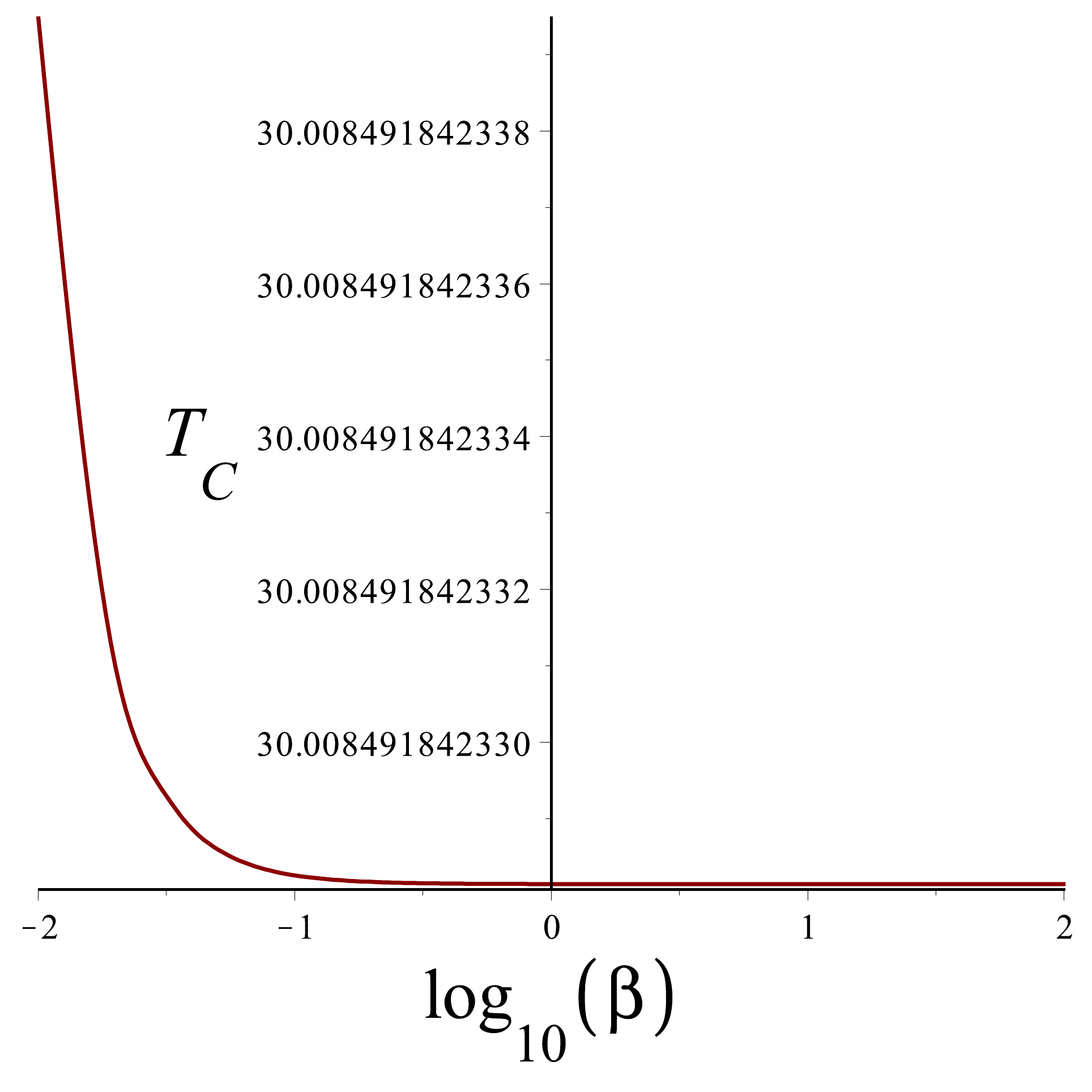}}\hspace{1.5cm}
\subfigure[]{\includegraphics[width=2.6in]{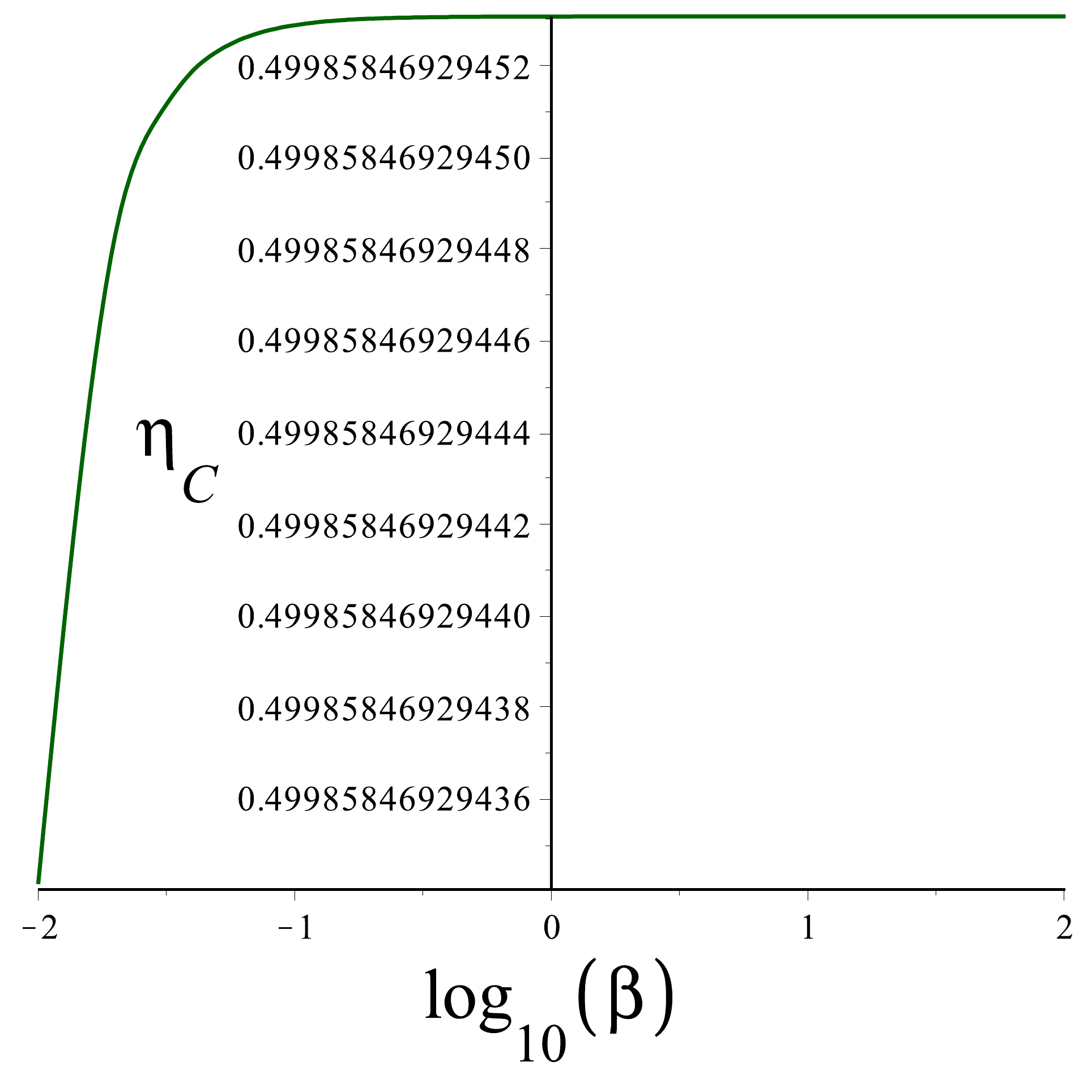} }
   \caption{\footnotesize  (a) The exact temperature $T_{\rm C}$ {\it vs.} $\log_{10}(\beta)$, in scheme~1. (b)  The  exact Carnot efficiency $\eta_{\rm C}$, {\it vs.} $\log_{10}(\beta)$, also in scheme~1. These quantities were computed using the exact equation of state. See text. (Here, we've chosen the values $p_1=5, p_4=3, T_1=50, T_2=60,$ and $q=0.1$. The same key features were observed for a range of sample values, including  even higher temperatures.)}   \label{fig:carnot-beta}
}
\end{figure}

 \begin{figure}[h]
{\centering
\subfigure[]{\includegraphics[width=2.6in]{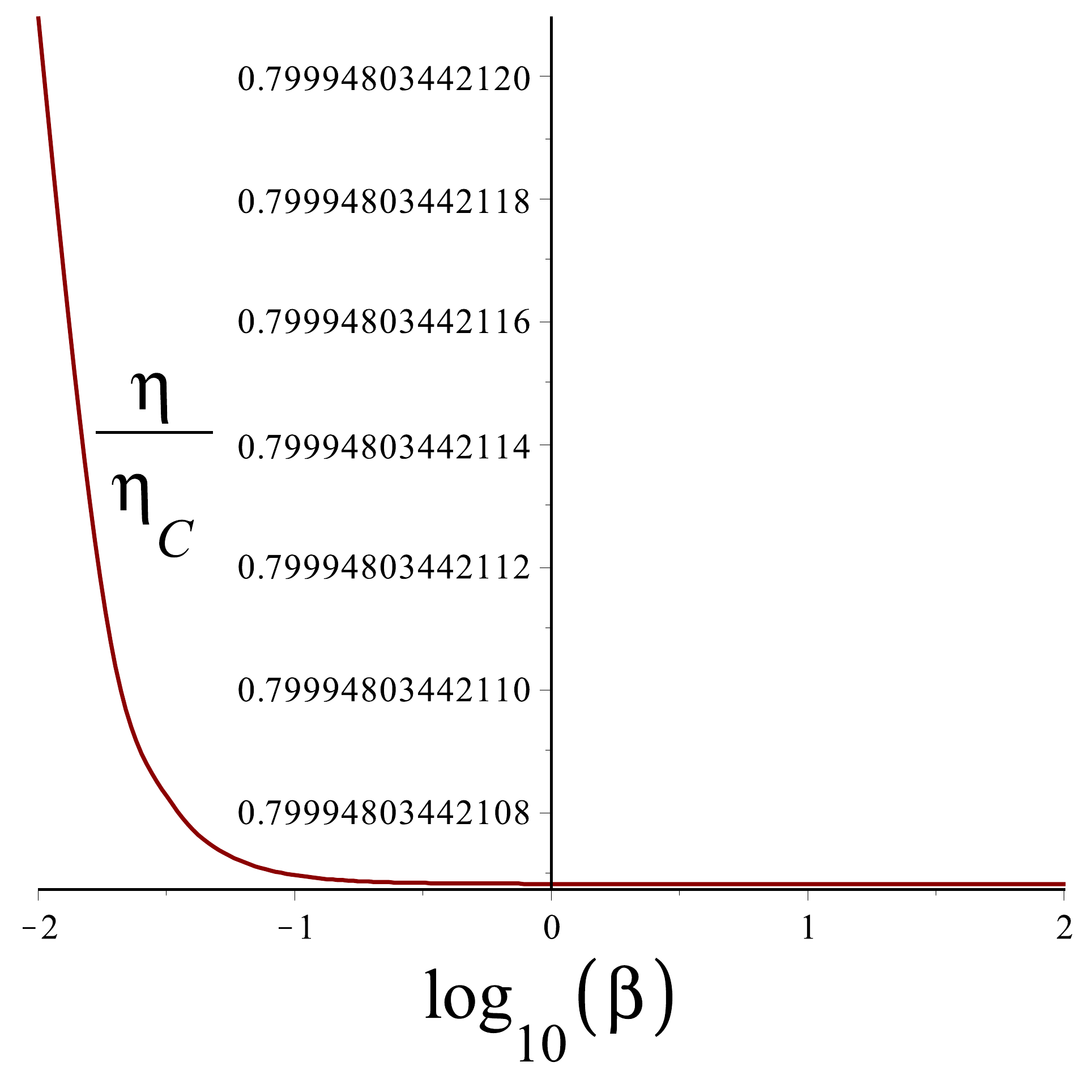}}\hspace{1.5cm}
\subfigure[]{\includegraphics[width=2.6in]{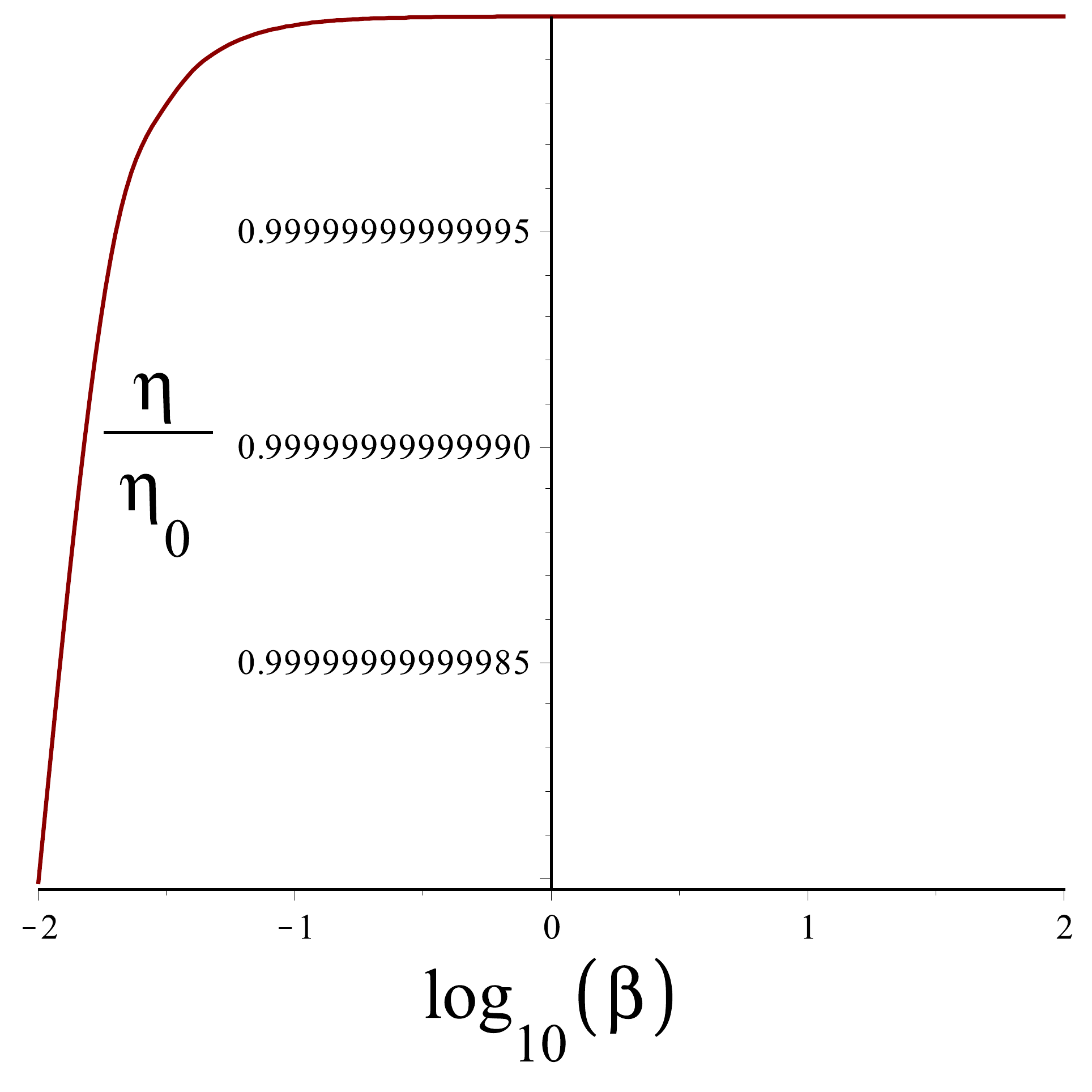} }
   \caption{\footnotesize  (a) The engine efficiency $\eta/\eta_{\rm C}$ {\it vs.} $\log_{10}(\beta)$, in scheme~1.   (b) The ratio $\eta/\eta_{ 0}$ {\it vs.} $\log_{10}(\beta)$ over the same range, also in scheme 1. (See the caption of figure~\ref{fig:carnot-beta} for the parameter values chosen.)}   \label{fig:efficiency-compare-1}
}
\end{figure}

Actually, at a given value of $\beta$ we can compare to an important additional  benchmark,  the Carnot efficiency $\eta^{\phantom{C}}_{\rm C}=1-T_C/T_H$, where $T_C$ and $T_H$ are, respectively, the lowest and highest temperatures our engine can attain. This  is the efficiency obtained  with a reversible heat engine operating between those  two temperatures.  Although we've specified $T_H\equiv T_2$, $\eta^{\phantom{C}}_{\rm C}$ changes with~$\beta$ since $T_C$ does: The equation of state must be used to determine $T_4\equiv T_C$.  We observe that as $\beta$ runs from $\infty$ toward smaller values $T_C$ increases slowly.  Correspondingly, the Carnot efficiency  decreases.   See figure~\ref{fig:carnot-beta} for the exact $T_C$ and $\eta^{\phantom{C}}_{\rm C}$ for a sample range $10^{-2}<\beta<10^2$. The Maxwell limit is to the right.  (This is analogous to what was seen in scheme~1 in ref.\cite{Johnson:2015ekr}.) As already remarked at the end of the last section, the dependence on $\beta$ is relatively weak, and for all quantities plotted in this section,  the variation is small over this wide range, with most of the change beginning late in the range at a ``turnaround region''   where, roughly,  $\beta\sim10^{-1}$.  All plots are against $\log_{10}(\beta)$   to better display the features. 

  Figure~\ref{fig:efficiency-compare-1}(a) displays the ratio $\eta/\eta^{\phantom{C}}_{\rm C}$ and  figure~\ref{fig:efficiency-compare-1}(b) shows that the ratio $\eta/\eta^{\phantom{9}}_0$, both plotted against $\log_{10}(\beta)$.  The Maxwell limit is to the right. In~\ref{fig:efficiency-compare-1}(a), it can be seen that the ratio grows slowly for  a while and then rises more rapidly  in the turnaround region. In~\ref{fig:efficiency-compare-1}(b) there is a very slow initial decline before the faster fall in the turnaround region. It is worth comparing this behaviour to that seen in the Gauss--Bonnet case for scheme~1,  as exhibited  in ref.\cite{Johnson:2015ekr}. We'll discuss the comparison further in  section~\ref{sec:closing}.

\subsection{Scheme 2}

\begin{figure}[h]
{\centering
\subfigure[]{\includegraphics[width=2.6in]{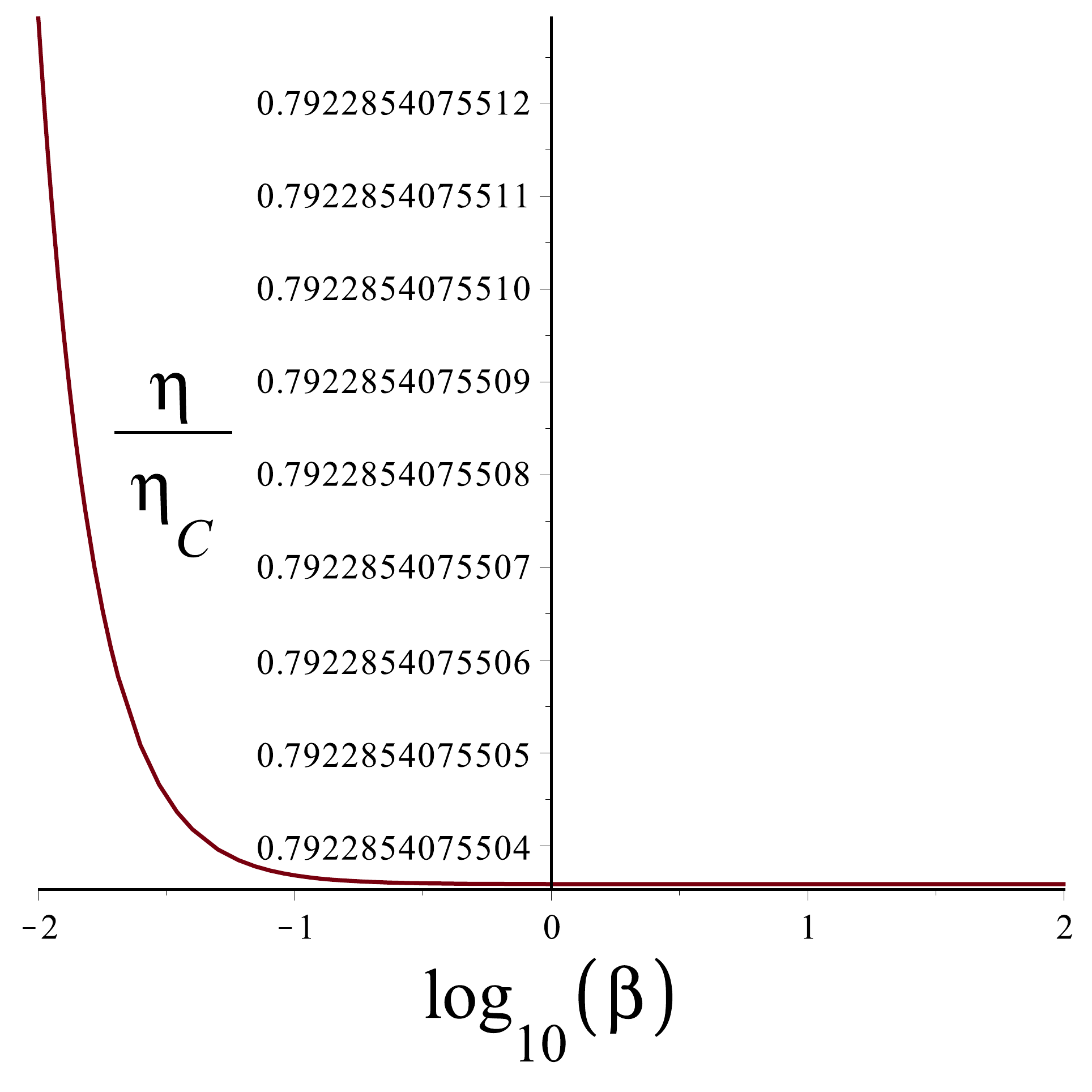} }\hspace{1.5cm}
\subfigure[]{\includegraphics[width=2.6in]{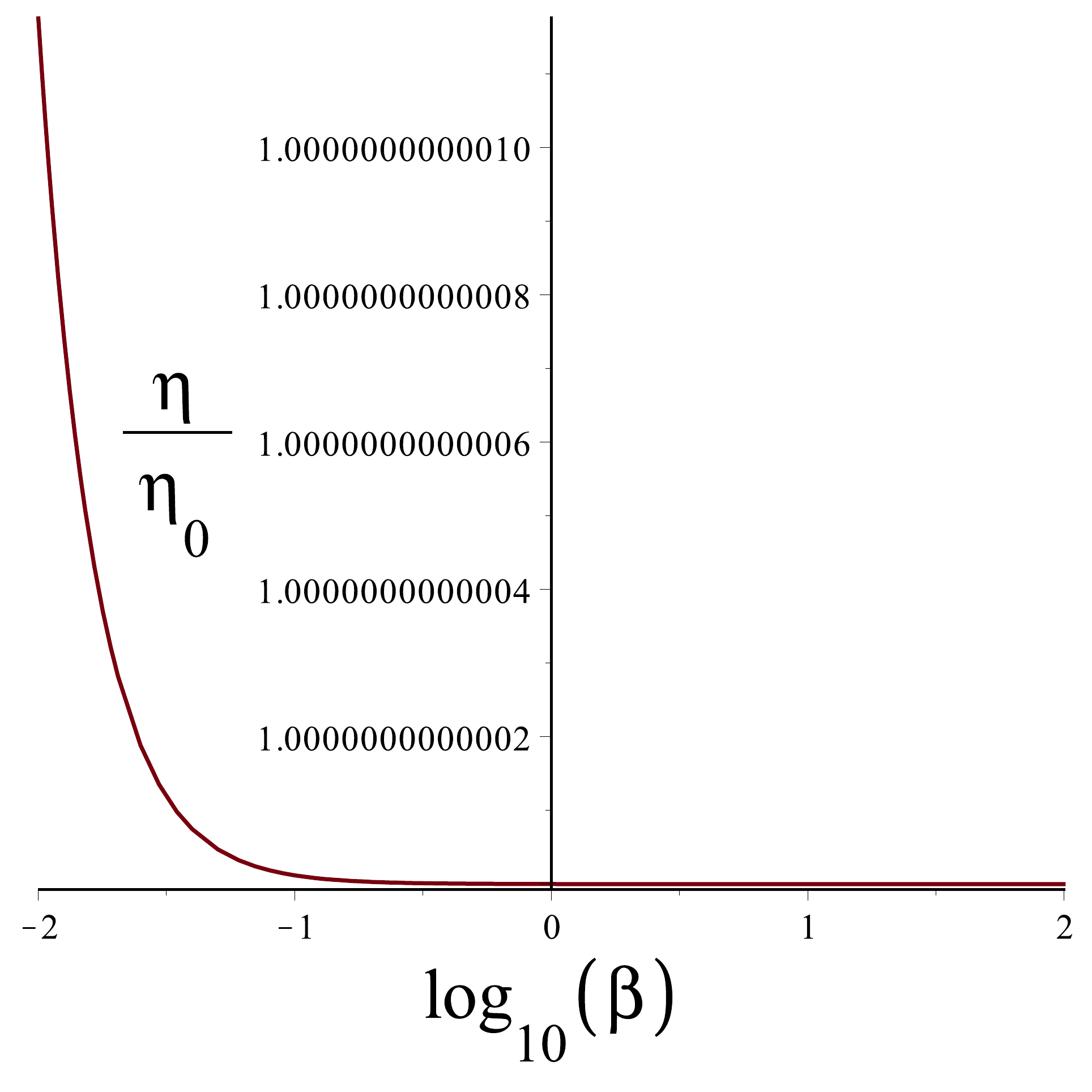} }

   \caption{\footnotesize  (a) The engine efficiency $\eta/\eta_{\rm C}$ {\it vs.} $\log_{10}(\beta)$, in scheme~2.  (b) The ratio $\eta/\eta_{ 0}$ {\it vs.} $\log_{10}(\beta)$ over the same range, also in scheme 2. (For these examples, $T_2\equiv T_H=60$, $V_2=33000$, $T_4\equiv T_C=30$, $V_4=15500$. The same key features were observed for a range of sample values, including  even higher temperatures.)}   \label{fig:efficiency-compare-2}
}
\end{figure}

In this scheme for our engine (again see figure~\ref{fig:cyclesb}) we instead specify the temperatures $(T_2,T_4)$, equivalent to specifying $(T_H,T_C)$, as well as the volumes $(V_2,V_4)$ (which also gives the pair $(V_3,V_1)$). Now the Carnot efficiency  $\eta^{\phantom{C}}_{\rm C}$ is fixed for all $\beta$. Instead, however, the pressures $p_1=p_2$ and $p_4=p_3$ must be determined using the equation of state, and so are now $\beta$--dependent. (We checked that the pressures in the engine remained physical over the range of $10^{-2}<\beta<10^2$, which is to be expected since we have fixed our highest and lowest temperatures to be far enough into the high temperature regime.) In figure~\ref{fig:efficiency-compare-2}, we again plot the ratios $\eta/\eta^{\phantom{C}}_{\rm C}$ and  $\eta/\eta^{\phantom{9}}_0$, against $\log_{10}(\beta)$. Again, the Maxwell limit is to the right. We'll discuss these results  further in  section~\ref{sec:closing}.

It is worth noting that the  $D=4$ case was explored explicitly as well, for each scheme, and the qualitative structure of  the results   was  found to be the same as for the $D=5$ case explored here, so no detailed results are reported from that case. The features (a slow change followed by the characteristic elbow or knee  in the turnaround region)  are a bit more pronounced since $\beta$ appears at slightly higher order: $O(T^{-5})$.  It is expected that higher $D$ will also work similarly.  


\section{Closing Remarks}
\label{sec:closing}
In ref.\cite{Johnson:2015ekr}, the corrections to the geometrical sector, from a Gauss--Bonnet term with coefficient $\alpha$, were studied for their effect on the efficiency of a heat engine in the same schemes~1 and~2 studied here. In the present study  we looked instead at corrections to the Maxwell sector, controlled by parameter $\beta$ in the Born--Infeld action, examining their effects on the heat engine efficiency. These two studies reveal features that are in sharp contrast to each other.

The most striking contrast is how weak the $\beta$--corrections are compared to the $\alpha$--corrections, as noted at the end of section
\ref{sec:high-temperature}, and as can be seen in all the figures in section~\ref{sec:twostudies}. The resulting magnitude of the variation in the efficiency with $\beta$ is of order $10^{-12}$. Even after  restrictions to the relatively narrow physical window allowed for $\alpha$, the variations there are many orders of magnitude greater\cite{Johnson:2015ekr}. (Our examples used in each case have the same values for the fixed parameters to allow this comparison.)

There is a much larger window of available values of $\beta$ to explore, while $\alpha$ is tightly constrained by certain physical requirements. This difference may ultimately be traceable to the fact that $\alpha$ controls just one finite set of correction terms in the action, while in contrast $\beta$ can explore a much richer range of effects starting with an infinite family of terms (in the sense of expanding the action~(\ref{eq:born-infeld}) around the $\beta\to\infty$ limit)  all the way to the  effects at small $\beta$ that generate the  turnaround region seen in section~\ref{sec:twostudies}. 

The large (but not infinite) $\beta$ regime is where we can best compare directly to the  small $\alpha$ regime that is the  physical window in  ref.\cite{Johnson:2015ekr}.  Here again, we see some contrasts. In scheme~1, the ratios $\eta/\eta^{\phantom{9}}_{\rm C}$ and   $\eta/\eta^{\phantom{9}}_0$ were seen to increase as the $\alpha$--corrections were turned on. The opposite is seen here for  the ratio $\eta/\eta^{\phantom{9}}_0$.  In scheme~2,   the behaviour of the ratios are of opposite character for the $\beta$ variations {\it vs.} the $\alpha$, variations, increasing  for the former case, and decreasing for the latter.

\section*{Acknowledgements}
 CVJ would like to thank the  US Department of Energy for support under grant DE-FG03-84ER-40168,  and Amelia for her support and patience.


\providecommand{\href}[2]{#2}\begingroup\raggedright\endgroup

\end{document}